\documentclass[11pt]{iopart}
\expandafter\let\csname equation*\endcsname=\relax
\expandafter\let\csname endequation*\endcsname=\relax
\usepackage{amsmath,amssymb,amsfonts,graphicx,textcomp}
\usepackage{tabularx}
\usepackage{color}
\usepackage[latin1]{inputenc}
\usepackage[T1]{fontenc}
\usepackage{verbatim}
\usepackage{graphicx}
\usepackage{bm}
\usepackage{lscape}
\usepackage{amsmath}

\begin{document}
\title{First principles characterization of the $P2_1ab$ ferroelectric phase of Bi$_2$WO$_6$}
\author{Hania Djani$^{1}$, Patrick Hermet$^{2}$ and Philippe Ghosez$^3$}
\address{
$^1$Centre de D\'eveloppement des Technologies Avanc\'ees,  cit\'e 20 aout 1956, BP. 17, Baba Hassen, Alger, Algeria  \\
$^2$Institut Charles Gerhardt Montpellier, UMR 5253 CNRS-UM2-ENSCM-UM1, Universit\'e Montpellier 2, Place E. Bataillon, 34095 Montpellier C\'edex 5, France\\ 
$^3$Physique Th\'eorique des Mat\'eriaux, Universit\'e de Li\`ege (B5), B-4000 Li\`ege, Belgium}
\ead{hdjani@ulg.ac.be, hdjani-aitaissa@cdta.dz}

%%%%%%%%%%%%%%%%%%%%%%%%%%%%%%%%%%%%%%%%%%%%%%%%%%%%%%%%%%%%%%%%%%
\begin{abstract}
The structural, dielectric, dynamical, elastic, piezoelectric  and  nonlinear optical (second-order susceptibility and Pockels tensors) properties of Bi$_2$WO$_6$ in its P2$_1ab$ ferroelectric ground state are determined using density functional theory.  The calculation of infrared and Raman spectra on single crystal allowed us to clarify the assignment of experimental phonon modes, considering the good agreement between the calculated and the experimental Raman spectra obtained on polycrystal. The calculation of the elastic constants confirms the elastic stability of the crystal and allow us to estimate the Young and shear moduli of polycrystalline samples. The piezoelectric constants have significant intrinsic values comparable to those of prototypical ABO$_3$ ferroelectrics. The electro-optic response is strongly dominated by the ionic contribution of transverse optic modes, yielding sizable Pockels coefficients around 9 pm/V along the polar direction. 
\end{abstract}
%%%%%%%%%%%%%%%%%%%%%%%%%%%%%%%%%%%%%%%%%%%%%%%%%%%%%%%%%%%%%%%%%%%%
\pacs{77.84.Bw, 77.80.-e, 63.20.dk, 77.22.Ej, 62.20.D-, 78.20.Jq, 78.30.Am}
\submitto{\JPCM}
\maketitle
%%%%%%%%%%%%%%%%%%%%%%%%%%
\section{Introduction}
%%%%%%%%%%%%%%%%%%%%%%%%%%

Bi$_2$WO$_6$ belongs to the family of Aurivillius phases, a kind of naturally layered structures in which  ferroelectric oxides can be observed \cite{1}. This family is described as perovskites blocks sandwiched between two fluorite-like layers. It has the (Bi$_2$O$_2$)$^{+2}$(A$_{m-1}$B$_{m}$O$_{3m+1}$)$^{-2}$ generic formula, where $m$ is an integer between 1 and 8 corresponding to the number of octahedra in the perovskites block~\cite{2}. Bi$_2$WO$_6$ is the end member of the Aurivillius family ($m=1$)  and possesses many interesting physical properties such as ferroelectricity associated to a large spontaneous polarization (P $\cong$ 50 $\mu$C/cm$^2$)  and a high Curie temperature (T$_c$=950$^\circ$C), piezoelectricity making it a potential alternative to BaTiO$_3$ and PbZr$_{1-x}$Ti$_{x}$O$_3$ solid solutions for applications at higher frequencies and temperatures~\cite{3}, high oxide ion conductivity~\cite{4}, and photocatalytic activity~\cite{5}. 

Bi$_2$WO$_6$ exhibits structural phase transitions with temperature, that  have been investigated in the literature~\cite{6,7}. Its ground state is an orthorhombic $P2_1ab$ (C$_{2v}^5$) ferroelectric structure that is related to a hypothetical paraelectric $I4/mmm$ (D$_{4h}^{17}$) phase  through the condensation of three unstable modes: a rotation of oxygen octahedra around the $c$-axis ($X_2^{+}$), a tilt of octahedra around the $a$-axis ($X_3^{+}$), and a polar zone-center distortion along the $a$-axis ($\Gamma_5^{-}$). An intermediate orthorhombic $B2cb$ (C$_{2v}^{17}$) ferroelectric structure, resulting from the condensation of $X_3^{+}$ and  $\Gamma_5^{-}$  modes only in the $I4/mmm$ paraelectric phase, is also observed above  $670^\circ C$. Experimentally, the $I4/mmm$ paraelectric phase is never reached at high temperature and a reconstructive ferroelectric-paraelectric phase transition takes place at 950$^\circ C$ to transform the material  from the $B2cb$ ferroelectric structure to the $A2/m$ (C$_{2h}^{3}$) paraelectric monoclinic phase. 

Surprisingly, although Bi$_2$WO$_6$ is generating interest for various applications, many of its physical properties have not been quantified yet. From the literature, we can only cite (i) the work of Yanovskii and Voronkova~\cite{8} reporting estimates of the dielectric, piezoelectric and optical properties of the ferroelectric $Aba2$ intermediate phase (conventional setting of $B2cb$) and (ii) the recent measurements by Maczka \textit{et al}.~\cite{9, 10} concerning the dynamical properties (Raman and infrared spectra) on the ferroelectric $Pca2_1$ ground state (conventional setting of $P2_1ab$).

In this paper, we use first-principles based methods to perform a comprehensive study of various physical properties of Bi$_2$WO$_6$  in its $P2_1ab$ ferroelectric phase. We calculate Raman and infrared spectra, providing benchmark theoretical data directly useful for the assignments of experimental spectra and clarifying the previously proposed assignments from Maczka \textit{et al}.~\cite{9, 10}. We also calculate the elastic and piezoelectric constants of this compound together with some nonlinear properties, such as: its second-order nonlinear optical susceptibilities and its Pockels coefficients. These quantities have not all been accurately quantified yet  in the literature while their estimation is mandatory to assess the interest of Aurivillius phases for potential applications.

%%%%%%%%%%%%%%%%%%%%%%%%%%%%%%%%%%%%%%%
\section{Technical details}
%%%%%%%%%%%%%%%%%%%%%%%%%%%%%%%%%%%%%%%%

First-principles calculations were performed within the density functional theory and the local density approximation (LDA) using the ABINIT package~\cite{11, 12, 13}. We used highly transferable Teter pseudopotentials~\cite{14}. Bi(5$d$, 6$s$, 6$p$), W(5$s$, 5$p$, 5$d$, 6$s$) and O(2$s$, 2$p$)-electrons were considered as valence states in the construction of the pseudopotentials. The electronic wavefunctions were expanded in plane-waves up to a kinetic energy cutoff of 55~Ha. Integrals over the Brillouin zone were approximated by sums over a 8$\times$8$\times$3 mesh of special $k$-points according to the Monkhorst-Pack scheme~\cite{15}. Relaxations of the lattice parameters and the atomic positions were performed using the Broyden-Fletcher-Goldfarb-Shanno algorithm \cite{16} until the maximum residual forces on the atoms and stress were less than 1$\times$10$^{-5}$ Ha/Bohr and 1$\times$10$^{-3}$ GPa, respectively.

Born effective charges, dielectric tensors and dynamical matrix (yielding the phonon frequencies and eigenvectors) were obtained within a variational approach to density functional perturbation theory~\cite{17}. The Raman susceptibilities, the second-order nonlinear optical susceptibilities and the Pockels tensor have been obtained within a nonlinear response formalism taking advantage of the 2$n+$1 theorem~\cite{18, 19}. The infrared absorption and Raman spectra were respectively calculated as described in Refs.~\cite{20} and~\cite{21}.

 \begin{figure}[t]
\centering\includegraphics[angle=0, scale=0.12]{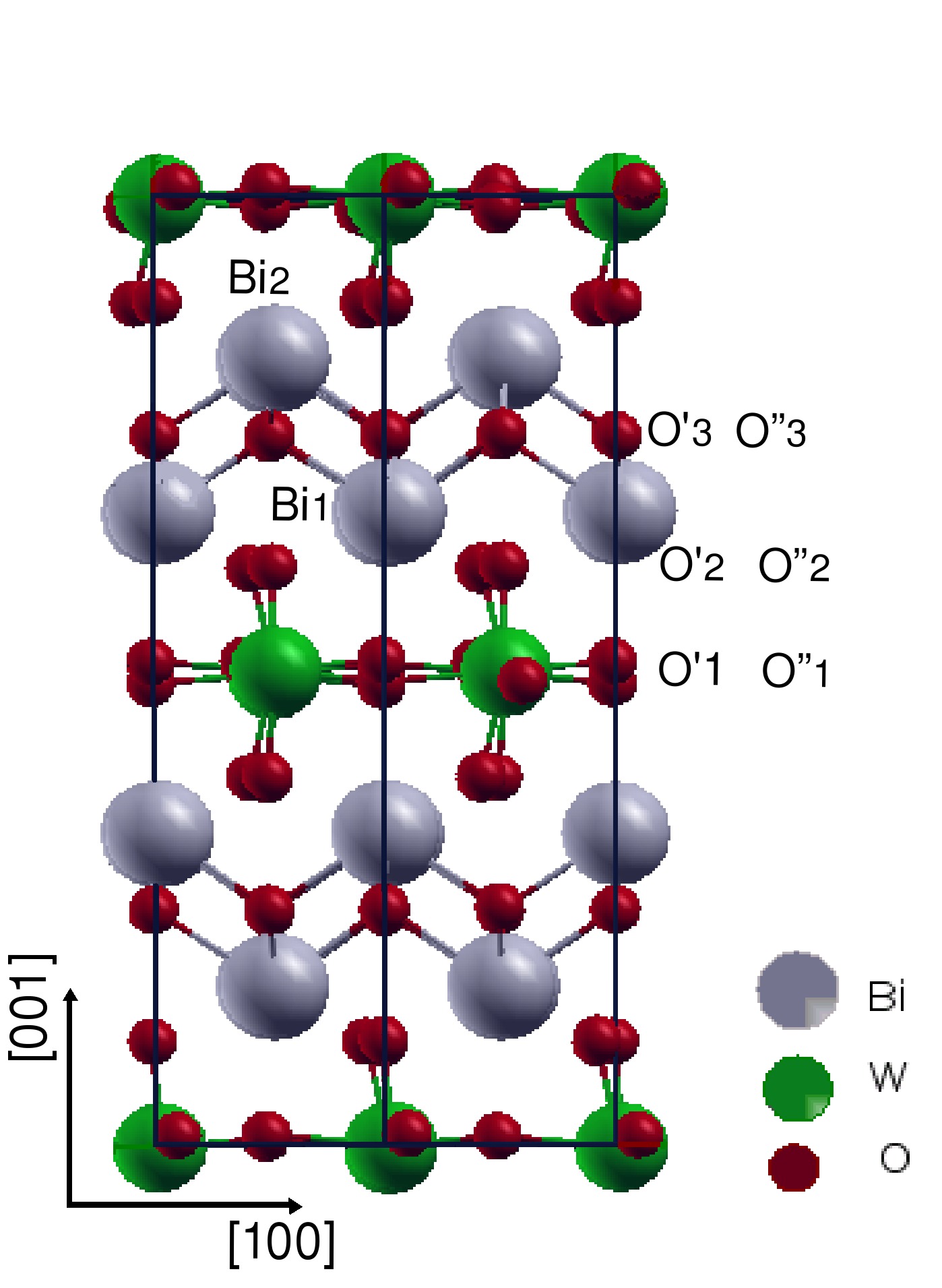}\\
\caption{\label{29struct} Primitive unit cell of Bi$_2$WO$_6$ in its $P2_1ab$ ferroelectric phase. [100] is the polar axis and [001] the stacking axis.}
\end{figure} 

Throughout this paper, we use the non-standard setting $P2_1ab$ to conform with the convention describing the ferroelectric phase. In this setting, the crystallographic $a$-axis is chosen as the polar axis, and the $c$-axis as the stacking axis that is orthogonal to the fluorite and perovskite layers (see Fig.~\ref{29struct}). All tensors will be reported in Cartesian coordinates where the orthogonal reference system ($x$, $y$, $z$) is chosen such as, $x$ is aligned along the polar axis and $z$ is aligned along the stacking axis.

In contrast, Maczka {\it et al.}~\cite{9} used the conventional $Pca2_1$ standard setting to investigate the polarized Raman spectra on single-crystal, where the polar axis is along the $c$-axis and the stacking direction along the $b$-axis. Thus, the comparison of their experimental results with ours was performed after the application of the matrix transformation from  $Pca2_1$ to $P2_1ab$: $$\begin{pmatrix} 0&1&0 \\ 0&0&1 \\ 1&0&0\end{pmatrix}.$$ 

In addition, we notice that comparisons with the experimental results of Yanovskii and Voronkova~\cite{8} (discussed later in this manuscript) have to be taken with caution since (i) we do not know exactly at which temperature they were measured,  (ii) they are {\it a priori} from the authors, associated to an $Aba2$ or $Pba2$ phase, experimentally recently proven  to be the intermediate phase within the temperature range 670$^\circ$C - 950$^\circ$C~\cite{6}, and (iii) the authors claim an accuracy of 10-20\% only. 

%%%%%%%%%%%%%%%%%%%%%%%%%%%%%%%%%%%%%%%
\section{$P2_1ab$ ground-state structure }
%%%%%%%%%%%%%%%%%%%%%%%%%%%%%%%%%%%%%%%

At room temperature, Bi$_2$WO$_6$ crystallizes in the $P2_1ab$ (C$_{2v}^5$) orthorhombic space group with the following experimental lattice constants: $a=$5.45~\AA, $b=$5.48~\AA\ and $c=$16.47~\AA~\cite{6}. These experimental lattice constants are slighly underestimated by  2\% in our calculations ($a=$5.30~\AA, $b=$5.32~\AA\ and $c=$16.17~\AA), as usually observed in LDA~\cite{7}. The nine independent atoms in this structure occupy the 4$a$ Wyckoff position, leading to a 36 atoms primitive unit cell ($Z=$4). The calculated atomic positions are given in Table~\ref{29param} and compared to the experimental ones measured at room temperature. In addition, to have a more quantitative comparison with the experimental data on the atomic distortions, we report in Table~\ref{overlap}, the projection of the atomic distortions from the hypothetical tetragonal paraelectric $I4/mmm$ phase to the orthorhombic ferroelectric $P2_1ab$ ground state, onto the phonon eigendisplacement vectors of the $I4/mmm$ phase. Only the most significant contributions are retained. These two tables show the good qualitative agreement between our predictions and the experimental data. However, we note from Table~\ref{overlap} that, due to the underestimation of the LDA volume, the amplitude of the distortions are globally underestimated in the relaxed structure. Similarly, the larger volume of the experimental structure leads to an increase of the  $\Gamma_5^{-}$ (29 $cm^{-1}$) stable polar mode and to a deacrease of the  $c$-axis rotation of oxygen octahedra ($X_2^{+}$ mode).

\begin{table}[h]
\begin{center}
\caption{\label{29param} Experimental~\cite{6} and calculated atomic positions of Bi$_2$WO$_6$ in its $P2_1ab$ ferroelectric phase.}
\begin{tabular}{ccccccccc}
\br
         &&\multicolumn{3}{c}{Calculated (0~K)}&&\multicolumn{3}{c}{Experimental (300~K)}\\
\cline{3-5}\cline{7-9}
Atoms    & & x/a & y/b & z/c && x/a & y/b& z/c\\
\hline
 Bi$_1$  & & -0.0109 & 0.5102 & 0.1698  && -0.0126 & 0.5191 & 0.1726\\
 Bi$_2$  & & -0.0036 & 0.4872 & -0.1698 && -0.0113 & 0.4839 & -0.1722\\
 W       & & 0.0000 & -0.0055 & 0.0000       &&  0.0000 & 0.0077& -0.0004 \\
 O'$_1$ & & 0.1679 & 0.7039 & -0.0108&& 0.2679 & 0.7015 & -0.0151\\
 O''$_1$  & & 0.2682 & 0.1951 & 0.0109  && 0.3342 & 0.2297 & 0.0159\\
 
 O'$_2$  & & 0.4532 & 0.5363 & 0.1114 && 0.5703 & 0.5603 & 0.1082\\
 O''$_2$ & & -0.0355 & 0.0491 & -0.1108 && 0.0854 & 0.0526 & -0.1076\\
 O'$_3$  & & 0.2464 & 0.2519 & 0.2483  && 0.2740 & 0.2403 & 0.2511\\
 O''$_3$ & & 0.2541 & 0.2468 & 0.7484 && 0.2728 & 0.2585 & 0.7485\\
\br
\end{tabular}
\end{center}
\end{table}

\begin{table}[h]
\begin{center}
\caption{\label{overlap} The contribution of the phonon modes of the hypothetical $I4/mmm$ phase (frequencies in brackets in cm$^{-1}$), to the distortion from $I4/mmm$ phase to calculated and experimental (300K) $P2_1ab$ ground state. Only the most significant contributions are retained. $A$ is the distortion amplitude (in Bohr) and $\alpha_i$ the cosine director, according to the conventions and normalization conditions defined in Ref. \cite{7}}.

\begin{tabular}{lccccccc}
\br
          && A   & & $\Gamma_5^{-}$ & $X_2^{+}$ & $X_3^{+}$ & $\Gamma_5^{-}$ \\
           &&      & & (198$i$) &(135$i$) & (104$i$) &(29) \\
\hline
Calculated (present) &&585  && 0.52  & 0.59  & 0.55  &0.04 \\
Experiment (300 K) \cite{6} &&772 && 0.56  &0.30   & 0.59   & 0.35  \\
\br
\end{tabular}
\end{center}
\end{table}

%%%%%%%%%%%%%%%%%%%%%%%%%%%%%%%%%%%%%%%%%
\section{Dielectric properties}
%%%%%%%%%%%%%%%%%%%%%%%%%%%%%%%%%%%%%%%%%

 \begingroup
\begin{table*}
\begin{center}
\caption{\label{born} Born effective charge tensors ($Z^\ast$) and their main values (between brackets) of the Bi$_2$WO$_6$ ferroelectric phase. The $Z^\ast$ of paraelectric phase are also reported between parenthesis for comparison.}

\tabcolsep 1.7pt
\small
\begin{tabular}{@{}cccccccccc}
\br
\\
$Z^*(Bi_1)$&$Z^*(Bi_2)$ &$Z^*(W)$  \\

$\begin{pmatrix} 4.94&0.0008&-0.36 \\ 0.02&4.76&-0.46 \\ 0.03&-0.37&4.42\end{pmatrix}$&$\begin{pmatrix} 4.81&0.18&-0.41 \nonumber\\ 0.12&4.86&0.64 \nonumber\\ -0.47&0.29&4.47\end{pmatrix}$&$\begin{pmatrix} 7.39&0.82&0.02 \\ -0.57&8.13&-0.13 \\ -0.13&-0.50&8.31\end{pmatrix}$\\
\\
\hspace{10pt}$\begin{bmatrix} 5.10&4.90&4.10\end{bmatrix}$&\hspace{10pt}$\begin{bmatrix}5.23&4.98&3.92\end{bmatrix}$&\hspace{10pt}$\begin{bmatrix} 8.42&8.04&7.35\end{bmatrix}$\\
\\
\hspace{10pt}$\begin{pmatrix} 5.00&5.00&4.29\end{pmatrix}$  & \hspace{10pt}$\begin{pmatrix} 5.00&5.00&4.29\end{pmatrix}$ &\hspace{10pt} $\begin{pmatrix} 11.09&11.09&9.32\end{pmatrix}$\\
\\
$Z^*(O'_1)$&$Z^*(O'_2)$ &$Z^*(O'_3)$  \\
$\begin{pmatrix} -3.59&2.32&-0.27 \\ 2.19&-3.56&0.09 \nonumber\\ -0.22&0.06&-1.42\end{pmatrix}$&$\begin{pmatrix} -2.34&-0.11&1.06 \nonumber\\ 0.16&-2.17&0.53 \nonumber\\ 0.52&0.62&-4.48\end{pmatrix}$& $\begin{pmatrix} -2.93&-0.2&-0.03 \nonumber\\ -0.002&-2.88&-0.04 \nonumber\\ -0.01&-0.21&-2.71\end{pmatrix}$\\
\\
\hspace{10pt}$\begin{bmatrix} -5.83&-1.49&-1.24\end{bmatrix}$&\hspace{10pt}$\begin{bmatrix} -5.68&-1.50&-1.22\end{bmatrix}$&\hspace{10pt}$\begin{bmatrix} -3.03&-2.84&-2.64\end{bmatrix}$\\
\\
\hspace{10pt} $\begin{pmatrix} -4.88&-4.88&-1.46\end{pmatrix}$  & \hspace{10pt}$\begin{pmatrix} -4.74&-2.70&-2.70\end{pmatrix}$ &\hspace{10pt} $\begin{pmatrix} -2.95&-2.95&-2.74\end{pmatrix}$\\
\\
$Z^*(O''_1)$&$Z^*(O''_2)$ &$Z^*(O''_3)$  \\
$\begin{pmatrix} -3.10&2.17&0.004 \\ 2.08&-3.93&-0.27 \\ 0.09&-0.28&-1.39\end{pmatrix}$ &$\begin{pmatrix} -2.27&-0.08&0.86 \\ -0.02&-2.28&-0.87 \\ 0.34&-0.89&-4.46\end{pmatrix}$&$\begin{pmatrix} -2.89&0.19&-0.05 \\ 0.003&-2.91&0.03 \\-0.04&0.20&-2.72\end{pmatrix}$\\
\\
\hspace{10pt}$\begin{bmatrix}-5.68&-1.50&-1.22\end{bmatrix}$&\hspace{10pt}$\begin{bmatrix} -4.88&-2.32&-1.80\end{bmatrix}$&\hspace{10pt}$\begin{bmatrix} -3.88&-2.70&-1.92\end{bmatrix}$\\
\\
\hspace{10pt}$\begin{pmatrix} -4.88&-4.88&-1.46\end{pmatrix}$  & \hspace{10pt}$\begin{pmatrix} -4.74&-2.70&-2.70\end{pmatrix}$ &\hspace{10pt} $\begin{pmatrix} -2.95&-2.95&-2.74\end{pmatrix}$
\\
\br
\end{tabular}
\end{center}
\end{table*}
\endgroup

Born effective charge tensors ($Z^\ast$) are dynamical quantities strongly influenced by dynamical changes of orbital hybridization induced by atomic displacements~\cite{22, 23}. $Z^*$ tensors have been calculated for the nine non-equivalent atoms of Bi$_2$WO$_6$  in its $P2_1ab$ ferroelectric phase and they are listed in Table~\ref{born}. The main values of these tensors have also been reported for a better comparison with Bi$_2$WO$_6$ nominal atomic charges ( +3, +6 and -2, for Bi, W and O atoms, respectively). We observe that the $Z^\ast$(W), $Z^\ast$(O$_1$) and $Z^\ast$(O$_2$) atoms are strongly anomalous compared to their nominal values. These anomalous Z* yield a strong spontaneous polarization of 48 $\mu$C/cm$^2$ in the $P2_1ab$ phase~\cite{7}.
These values can be linked to hybridization between the 2$p$-orbitals of O$_1$ and O$_2$  types of oxygens and the 5$d$-orbitals of $W$-atoms, as reported for the ABO$_3$ prototypical ferroelectrics~\cite{24, 25}.

We observe also sizable anomalous Z* on $Bi$ and $O_3$ atoms; in this case however, the lone pair of Bi could play an active role beyond purely hybridization effects. We note that the off-diagonal terms of $Z^\ast$ are very small except for $O_1$-atoms in the basal plane of the crystal.

The {\it optical} dielectric tensors of the ferroelectric and the paraelectric phases of Bi$_2$WO$_6$  are reported in Table~\ref{optdielc}. In our theoretical framework, they correspond to the purely electronic response to a static field. It is well known that DFT (LDA) usually overestimates the absolute value of $\varepsilon_\infty$ with respect to the experiment. This problem is linked to the underestimation of the electronic bandgap and the lack of polarization dependence of local (LDA) exchange-correlation functionals~\cite{26}. To overcome this problem, it is a common practice to apply the so-called ``\textit{scissors correction}''~\cite{27}, in which we use an empirical rigid shift of the conduction bands to adjust the LDA band-gap to the experimental value. By comparing our calculated band-gap value ($E_g^{calc}= $1.9 eV) with the experimental one ($E_g^{exp}= $2.6 eV~\cite{28}), the scissors correction is fixed to 0.7 eV. As expected, this correction slightly decreases the values of the optical dielectric tensor (see Table~\ref{optdielc}).

\begin{table}[tb]
\begin{center}
\caption{\label{optdielc} The optical dielectric tensor components ($\varepsilon^\infty_{ij}$) and related refractive indices ($n_{i}=\sqrt{\varepsilon^\infty_{ii}}$) of the hypothetical paraelectric ($I4/mmm$)  and ground-state ferroelectric ($P2_1ab$) phases of Bi$_2$WO$_6$. Values after a scissors correction fixed to 0.7 eV are also given for the ferroelectric phase.}
\begin{tabular}{lccccccccc}
\br
Phase  &Method & &$\varepsilon^\infty_{11}$ &$\varepsilon^\infty_{22}$&$\varepsilon^\infty_{33}$  && $n_{1}$ & $n_{2}$ & $n_{3}$  \\
\hline
$I4/mmm$ &LDA &&7.60 & 7.60&6.61 &&2.76  &2.76  &2.57  \\
$P2_1ab$ &LDA &&6.60 & 6.69&6.61 &&2.57  &2.59  &2.57     \\
$P2_1ab$ &LDA+SCI &&6.09 & 6.16&6.09 &&2.47  &2.48  &2.47    \\
\br
\end{tabular}
\end{center}
\end{table} 

The optical dielectric tensor components and related refractive indices in the ferroelectric phase have values smaller than those observed in the paraelectric phase. The refractive indices in the $P2_1ab$ phase after the scissors correction have amplitudes comparable to what was measured by Yanovskii and Voronkova ($n \approx 2.5$)~\cite{8}. We also predict a weak negative birefringence (reduction of $n$ along the polar axis), but with a birefringence amplitude one order of magnitude smaller than estimated experimentally. 
 
The {\it static} (or low-frequency) dielectric tensor of the $P2_1ab$ ferroelectric phase can also be estimated theoretically. The fixed-strain relaxed-ion dielectric tensor, $\varepsilon^{(\eta)}$, can be obtained by adding to the purely electronic response ($\varepsilon^{\infty}$)  the contribution coming from the response of the ions to the electric field. To estimate this last contribution, one can use a model that assimilates the solid to a system of undamped harmonic oscillators. Doing so, $\varepsilon^{(\eta)}$ appears as the sum of an electronic contribution ($\varepsilon^{\infty}$) and a contribution arising from each individual polar phonon mode ($\varepsilon_m^{ph}$) such as~\cite{17}:
\begin{equation}
\label{epsilon_0}
 \varepsilon_{\alpha\beta}^{(\eta)}=\varepsilon_{\alpha\beta}^{\infty}+\sum_m \varepsilon_{\alpha\beta,m}^{ph} = \varepsilon_{\alpha\beta}^{\infty}+\frac{4\pi}{\Omega_0}\sum_m \frac{S_{\alpha\beta}(m)}{\omega_m^2},
\end{equation}
where the sum runs over all polar modes $m$, $\Omega_0$ is the unit cell volume and $S$ is the infrared oscillator strength. The contributions arising from the ionic relaxation can be estimated from the phonon calculation that will be further discussed in the next Section. The results are presented in Table~\ref{statdielect}.  It appears from the mode by mode decomposition that  $\varepsilon_{11}^{(\eta)}$ is dominated by the A$_1$-mode calculated at 62~cm$^{-1}$ . Similarly, two B$_1$-modes calculated at 59 and 212~cm$^{-1}$ dominate $\varepsilon_{22}^{(\eta)}$. In contrast, no clear contribution of the B$_2$-modes governs  $\varepsilon_{33}^{(\eta)}$.

\begin{table}[tb]
\begin{center}
\caption{\label{statdielect} The fixed-strain relaxed ion ($\varepsilon^{(\eta)}$) and relaxed-stress relaxed-ion ($\varepsilon^{(\sigma})$) static dielectric tensors of Bi$_2$WO$_6$ in its $P2_1ab$ phase (including $\varepsilon^{(\infty)}$ with ``scissors'' correction) . Theoretical results are compared to experimental data ($\varepsilon^{exp}$) from Yanovskii {\it et al.}~\cite{8}. }
\begin{tabular}{lcccccccc}
\br
Index & &$\varepsilon^{(\eta)}$& &$\varepsilon^{(\sigma)}$ &&$\varepsilon^{exp}$       \\
\hline
11  & &39 &&50 & &$\approx$ 100     \\
22  &  &43 &&60 &  &$ \approx$ 60                \\
33  &  &11 &&12  &   &$\approx$ 60\\
\br
\end{tabular}
\end{center}
\end{table}  
 
The relaxed-stress relaxed-ion dielectric tensor, $\varepsilon^{(\sigma)}$,  can also be computed by adding to $\varepsilon^{(\eta)}$ the contribution associated to the strain relaxation as described in Ref.~\cite{29}. We observe in Table~\ref{statdielect} that the strain relaxation has a strong effect on the static dielectric constant in the directions perpendicular to the stacking axis. 

Typically, $\varepsilon^{(\sigma)}$ should be compared to \textit{ac} dielectric measurements at frequencies much less than sample resonances while $\varepsilon^{(\eta)}$  would correspond to frequencies much higher than the sample resonance but much less than phonon frequencies. We notice that our theoretical prediction significantly underestimate the experimental value (at 1 kHz) along the polar direction~\cite{8}. This could be linked to the fact that our LDA calculations slightly underestimate the lattice constants,  which in turn could harden some relevant modes. As it is discussed in the next Section, the A$_1$-mode dominating $\varepsilon_{11}^0$ is nevertheless predicted at 62~cm$^{-1}$ in close agreement with experimental data at low temperature (59~cm$^{-1}$). A possible explanation is that, although the ferroelectric transition is at very high temperature and reconstructive,  the frequency of this dominating A$_1$-mode is temperature dependent and much softer at the experimental measurement temperature than in the zero Kelvin limit. Much more surprising is the large experimental value $\varepsilon_{33}^{(exp)}$ reported along the stacking axis, and the absence of experimental anisotropy between $\varepsilon_{22}^{(exp)}$ and $\varepsilon_{33}^{(exp)}$~\cite{8}.

%%%%%%%%%%%%%%%%%%%%%%%%%%%%%%%%%%%%%%%%%%%%%%%%%%%%%%%%%%%%%%%%%%
\section{Optical phonon modes at the zone-center}
%%%%%%%%%%%%%%%%%%%%%%%%%%%%%%%%%%%%%%%%%%%%%%%%%%%%%%%%%%%%%%%%%%

Bi$_2$WO$_6$ in its ferroelectric $P2_1ab$ structure belongs to the $C_{2v}$ point group. The zone-center optical phonons can be therefore classified according to the irreducible representations of this group as: $\Gamma_{opt}$ = 26 A$_1$ $\oplus$ 26 B$_1$ $\oplus$ 26 B$_2$ $\oplus$ 27 A$_2$. The first three irreducible representations are both infrared and Raman active. They are polarized along $x$, $y$ and $z$-direction, respectively. The last representation (A$_2$) is only Raman active. The form of the Raman susceptibility tensors in the $P2_1ab$ non-standard setting is reported in Table~\ref{ramantensor}. Close to the $\Gamma$-point, the macroscopic electric field splits the polar active modes into transverse (TO) and longitudinal (LO) modes. Table~\ref{assign} compares our calculated TO and LO phonon frequencies classified according to their symmetry with the corresponding experimental frequencies measured on single crystal by  Maczka {\it et al.}~\cite{9}. An overall acceptable agreement is observed.

\begingroup
 %\squeezetable
\begin{table}[h!]
\caption{\label{ramantensor}Raman susceptibility tensors of the A$_1$, B$_1$,A$_1$ and B$_2$ modes in the non-standard setting $P2_1ab$ space group.}
 %\begin{indented}
\centering 
   %\tiny
\begin{tabular}{@{}ccccccc}
\br
 $A_{1}(x)$=$\begin{pmatrix} a&0&0 \\ 0&b&0 \\ 0&0&c\end{pmatrix}$&$B_{1}(y)$=$\begin{pmatrix} 0&d&0 \nonumber\\ d&0&0 \nonumber\\ 0&0&0\end{pmatrix}$\\
\\
$A_{2}$=$\begin{pmatrix}0&0&0 \\ 0&0&f \\ 0&f&0\end{pmatrix}$&$B_{2}(z)$=$\begin{pmatrix} 0&0&e \nonumber\\ 0&0&0 \nonumber\\ e&0&0\end{pmatrix}$\\
\br
\end{tabular}
 %\end{indented}
\end{table}
\endgroup

\begin{landscape}
\begin{table}[]
\caption{\label{assign} Calculated and experimental frequencies (in $cm^{-1}$) of the zone-center optical phonon modes of a Bi$_2$WO$_6$ single crystal in its $P2_1ab$ phase. Experimental frequencies are obtained using infrared (IR) and Raman (Ra) spectroscopies by Maczka {\it et al.}~\cite{9}.} 
\small
\begin{center}
\begin{tabular}{ccccccccccccccccccccccccccc}
\br
\multicolumn{3}{c}{A$_1$(TO)}&&\multicolumn{3}{c}{A$_1$(LO)}&&\multicolumn{2}{c}{B$_2$(TO)}&&\multicolumn{1}{c}{B$_2$(LO)}&&\multicolumn{2}{c}{A$_2$(TO)}&&\multicolumn{3}{c}{B$_1$(TO)}&&\multicolumn{2}{c}{B$_1$(LO)}  \\
\cline{1-3}\cline{5-7}\cline{9-10} \cline{12-12}\cline{14-15}\cline{17-19}\cline{21-22}
IR  & Ra  & DFT   &&   IR  & Ra  & DFT   &&    Ra & DFT   &&DFT&&    Ra& DFT   &&   IR  & Ra  & DFT     &&   IR & DFT   \\
\br
  - &  36     &   35      &&    -   &  36    &   35      &&    65     &   50     && 50 &&   56    &   53       &&    -   &   -    &  36        &&   -   &  36        \\
 - &   59    &   62     &&    -   &  59  &    72    &&    78     &   75      &&76 &&    76     &   68       &&    -   &   58    &  59        &&   -   &  69        \\
 -  &  68    &   74     &&    -   &  69  &     76      &&   -      &   81     &&82 &&     -     &   82       &&    -   &   71   &  87        &&   -   &  87       \\
 -  &  77       &   84     &&    -   &  77     &   85      &&   -      &   82     &&82  &&    -     &   84       &&   -   &   89   &  103        &&   -   &  103        \\
 -  &  98       &   96     &&    -   &  98     &   96      &&    94       &   100     &&100 &&    -     &   90       &&   -   &   98   &  112        &&   -   &  113        \\
 -  &  -       &   114    &&    -  &  -      &   115      &&    -      &   103    &&120 &&    100    &   107       &&    -  &   150   &  165        &&   -   &  166        \\
 -  &  140       &   166      &&    -   &  -     &   175      &&    150      &   160      &&162  &&    -   &   108      &&    -   &   -  &  166        &&   -   &  167        \\
 -  & 150       &   175   &&    -   &  154    &   178      &&    -    &   166   &&167 &&    -    &  122       &&    -   &   182 &  174        &&   -   &  175        \\
 -  &  183       &   180     &&    -   &  184     &   181      &&   -      &   168      &&169  &&    -    &   172  &&    -   &   - &  191        &&   -   &  191        \\
 -  &  206       &   185    &&    -   &  209      &   187      &&    210     &   185      &&185 &&    218   &   192 &&    -   & 206  &  212        &&   -   &  265        \\
 -  &  258       &   254    &&    -   &  -     &   261      &&   -      &   267      &&268 &&     257   &   232 &&    -   &   258  &  267        &&   -   &  270        \\
 -  &  -       &   261  &&    -   &  -     &   266      &&   -      &   275     &&275  &&    282    &   286   &&    -   &   285  &  289       &&   -   &  296        \\
 -  &  -       &   267    &&    -   &  -     &   271      &&   285     &   291      &&291  &&    306    &   300  &&    -   & 307 &  310        &&   -   &  325        \\
 -  &  -       &  274     &&    -   &  -      &   282      &&     303     &   296      &&297 &&    332   &   312 &&    -   &  343 &  331        &&   -   &  334        \\
 -  &  284       &   282      &&    -   &  286      &   299      &&   -     &   299      && 301 &&    -   &   334  &&    -   &  -  &  334        &&   -   &  340        \\
 -  &  307       &   309      &&    -   &  335     &   337      &&    335     &   309      &&309 &&   -  &   358 &&    -   &  - &  340        &&   -   &  412        \\
 -  &  338       &   353    &&       &  -   &   358      &&   -    &   365      &&421 &&     -   &   359 &&    422   &   - &  417        &&   424   &  420       \\
 -  &  369       &   358    &&       &  370     &   385      &&   -      &   424      && 426 &&    420   &   420 &&    -   &   - &  429        &&   -   &  433        \\
 -  &  -       &   386  &&    -   &  -     &   389      &&   -      &   452      &&491 &&     -   &   432 &&    444   &   -   &  447        &&   -   & 448        \\
 -  &  -       &   389    &&    -   &  -     &   421      &&   -     &   491      && 492 &&   -  &   508 &&    -   &   -  & 450        &&   471   &  469        \\
 420  &  418       &   421      &&    -   &  -      &   427      &&   -    &   518      &&584 &&    -  &   509 &&    -   &   - &  543        &&   -   &  544        \\
 -  &  -       &   428  &&    458   &  456     &   459      &&   -     &   585      &&586  &&    -   &   546       &&    596   &   - & 557        &&   599   &  558       \\
 734  &  728       &   713      &&    753   &  -      &   774      &&    -    &   694      && 713&&   598  &   560  &&705   &   -  &  634        &&   -   &  636       \\
 -  &  -       &   776  &&    -   &  -      &   793      &&    704    &   713 &&776 &&    -  &   635  &&    -   &   688   & 686        &&   745   &  707       \\
 803  &  794       &   793      &&    803   &  798      &   794      &&   -      &   793      &&793 &&     -  &   686 && 757   &  795 & 707        &&   824   &  803      \\
 829  &  818  &   813     &&    843  &  845      &   832      &&    844  &   811      &&814 &&    703 &   707      &&    -   &   818  & 997       &&   -   &  998       \\
  &      &     &        &&       &     &     &         &&     &         &&-& 997  &      &&    &     &   &        &&     &         \\
\br
\end{tabular}
\end{center}
\end{table}
\end{landscape}

\begin{figure}[t]
\centering\includegraphics[angle=0, scale=2.5]{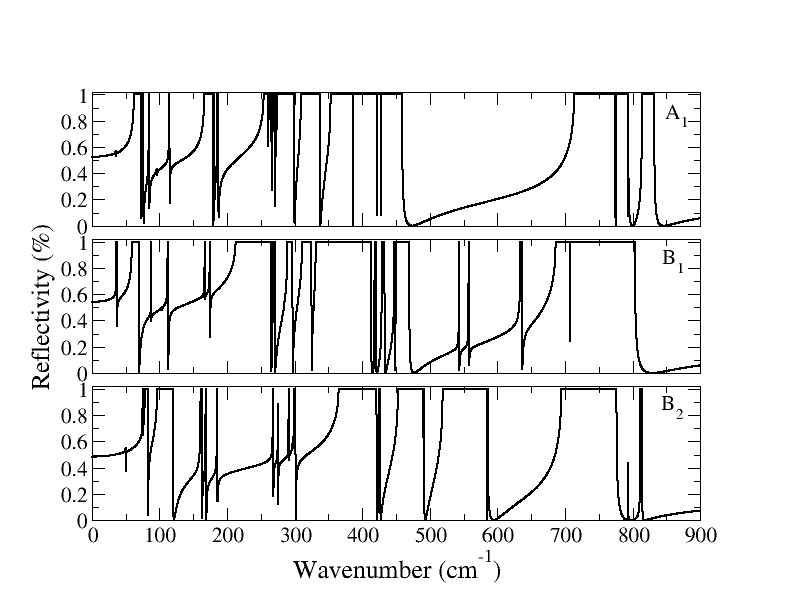}\\
\caption{\label{IR} Calculated infrared reflectivity spectra of Bi$_2$WO$_6$ in its $P21ab$ phase: A$_1$-modes (E//x), B$_1$-modes (E//y) and  B$_2$-modes (E//z).}
\end{figure}

The analysis of the infrared reflectivity spectra allows to quantify the LO--TO splitting strength. These spectra are calculated at normal incidence for the three polar irreducible representations and are displayed in Fig.~\ref{IR}. The reflectivity saturates to the unity because our formalism neglects  the damping of the phonon modes. In agreement with the experimental data~\cite{9} reported in Table~\ref{assign}, we observe a non negligible LO--TO splitting of the B$_1$-modes around 200, 340 and 700 cm$^{-1}$, while this splitting is located in the  250--430 cm$^{-1}$ range  and around 700 cm$^{-1}$ for the A$_1$-modes. The splitting of the B$_2$-modes has not been measured experimentally but our calculations show that it is significant  in the 360--500 cm$^{-1}$ range  and around 700 cm$^{-1}$.

\begin{figure}[t]
\centering\includegraphics[angle=0, scale=2.5]{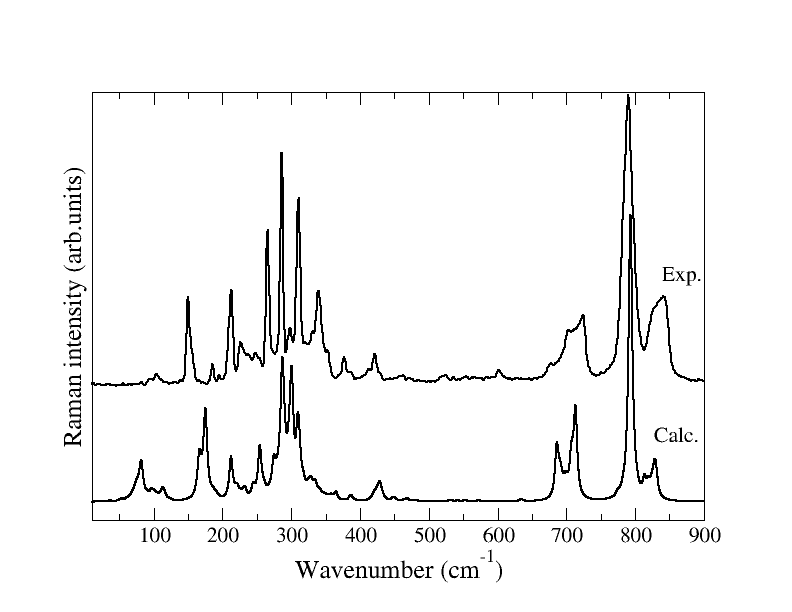}\\
%\centering\includegraphics[angle=0, scale=0.45]{ramancalcpoly.png}\\
\caption{\label{ramanpoly} Calculated and experimental Raman spectra of Bi$_2$WO$_6$  polycrystals. Spectra are normalized on the strong line centered at 790~cm$^{-1}$. The experimental spectrum is recorded at 7~K by Maczka {\it et al.}~\cite{10}. The calculated spectrum is displayed using a Lorentzian line shape with a constant linewidth fixed at 3~cm$^{-1}$. }
\end{figure}

Figure~\ref{ramanpoly} compares the calculated unpolarized Raman spectra of Bi$_2$WO$_6$ polycrystalline powder and the experimental one recorder by Maczka {\it et al.}~\cite{10} at 7~K, in the 10--900~cm$^{-1}$ range. 

\begin{figure}[h]
\centering\includegraphics[angle=0, scale=0.30]{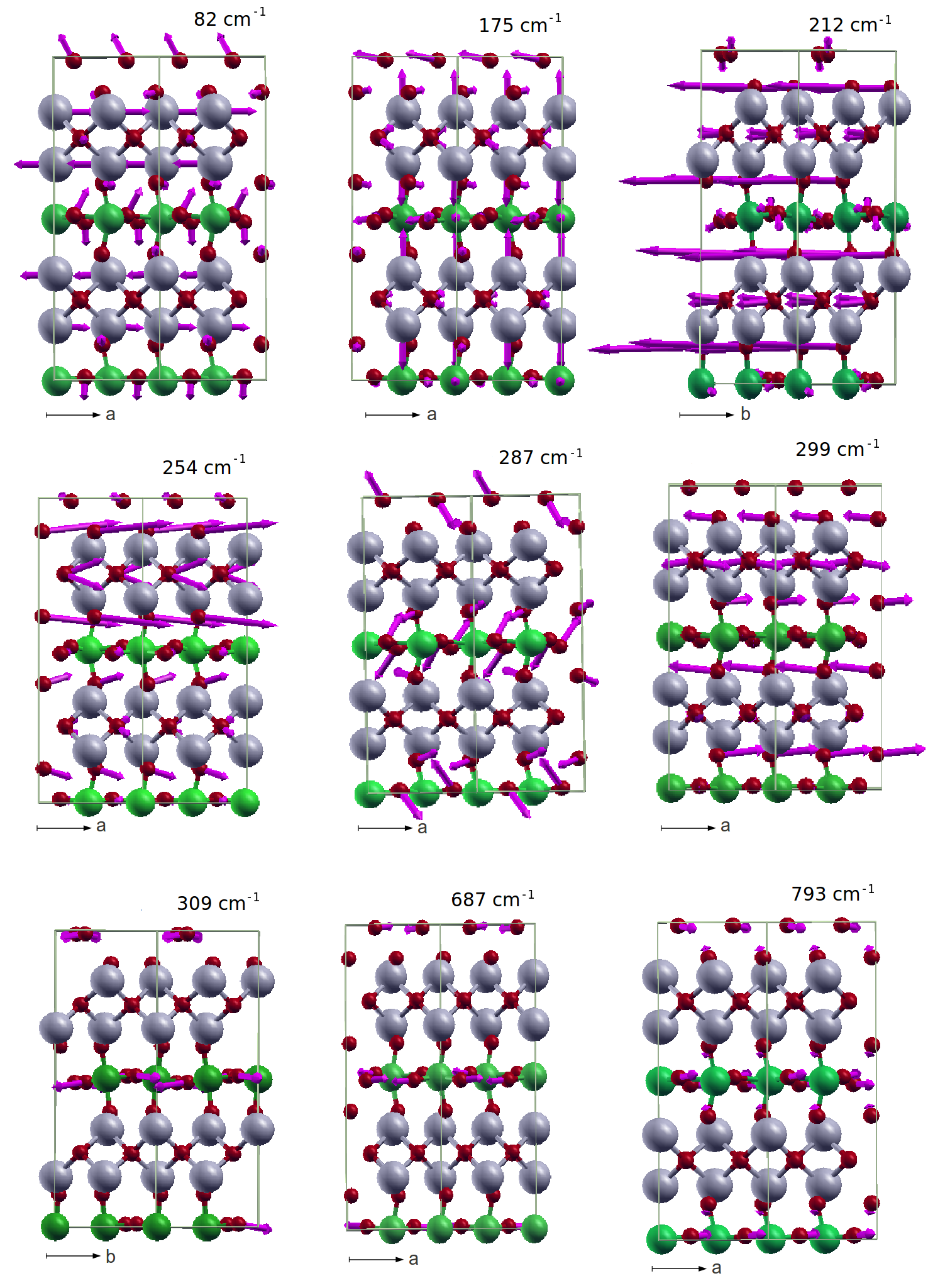}\\
\caption{\label{atomdispla} Eigendisplacement vectors of some calculated Raman lines of Bi$_2$WO$_6$. Arrows are proportional to the amplitude of the atomic motion.}
\end{figure}

The calculated average spectrum is performed assuming a quasi-continuous and random distribution of crystallites orientations. In practice, this is done by evaluating the Raman tensor components for an arbitrary orientation in space using Euler angles~\cite{30,32}. The experimental Raman spectrum is dominated by 14 lines centered at: 150, 213, 226, 265, 286, 311, 340, 377, 421, 703, 724, 790, 827 and 842~cm$^{-1}$. Their frequency position and their relative intensity are well reproduced by the calculation, except below 250~cm$^{-1}$ where the intensities are overestimated. The assignment of the low-frequency phonon modes (range below 550~cm$^{-1}$) remains still experimentally unexplored. We assigned them thanks to the analysis of their eigendisplacement vectors, obtained from the diagonalization of the dynamical matrix. This assignment is reported in Table~\ref{ass}, while Fig.~\ref{atomdispla} displays the eigendisplacement vectors of some of them. In Table~\ref{ass}, a mode is assigned as TO+LO when its TO--LO splitting is equal or less than 5~cm$^{-1}$. Similarly, we assigned the high-frequency phonon modes (550--850~cm$^{-1}$ range) and succeed to clarify the previous assignments obtained by Maczka {\it et al.}~\cite{9} using polarized Raman spectroscopy on single-crystal (see Table~\ref{ass}). 

\begin{figure}[t]
\centering\includegraphics[angle=0, scale=2.5]{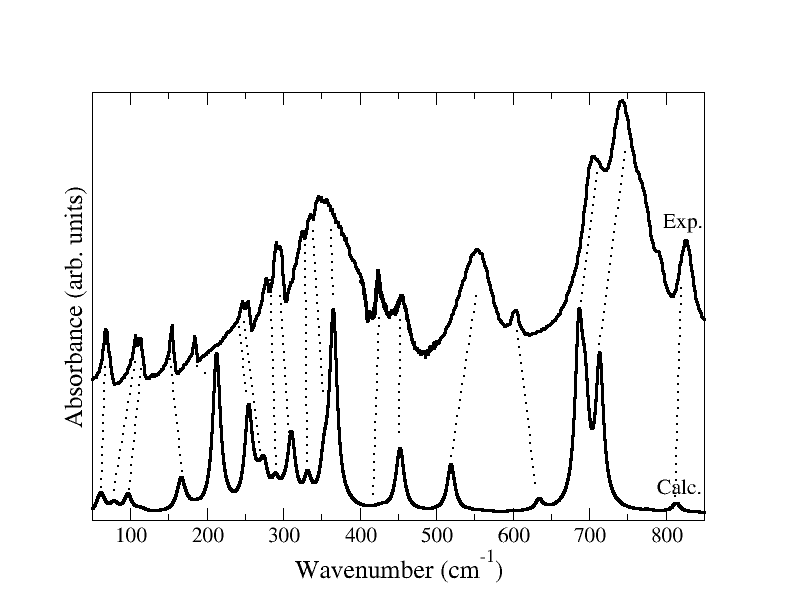}\\
\caption{\label{IRpoly} Calculated and experimental infrared spectra of Bi$_2$WO$_6$  polycrystals. The experimental spectrum is recorded at 7~K by Maczka {\it et al.}~\cite{10}.
The calculated spectrum is displayed using a Lorentzian line shape with a constant linewidth fixed at 6~cm$^{-1}$. The dashed lines show the assignment of the main experimental bands to the corresponding calculated ones.}
\end{figure}

The infrared absorption spectrum on Bi$_2$WO$_6$ polycrystalline powder was also calculated to complete the knowledge of its zone-center phonon modes. It is displayed in Fig.~\ref{IRpoly}, together with the experimental recorded at 7~K. The latter is obtained by grinding the  powder with a mulling agent, in contrast to the experimental Raman spectrum~\cite{9,10}. We observe that the agreement experiment--calculation is much less satisfactory than for the Raman. It is quite surprising because the calculation of infrared spectra is much simpler than the calculation of Raman spectra. Thus, this poor agreement could be related to multiple scattering within the particles due to their size larger than the infrared wavelength (Mie scattering). Nevertheless, the assignment of the main experimental bands to the corresponding calculated ones is still possible as highlighted by the dashed lines in the figure.

\begin{landscape}
\begin{table}[t]
\caption{\label{ass} Assignment of the most dominant Raman lines of a Bi$_2$WO$_6$ polycrystal in its $P2_1ab$ ferroelectric phase. Notation: $\nu_a =$ antisymmetric stretching, $\nu=$ stretching, $\tau =$  twisting, $\delta =$ bending, $R_u = $ rotation around $u$-axis , $t_u =$ translation along $u$-axis.}
\begin{center}
\begin{tabular}{ccccc}
\br
\multicolumn{2}{c}{Freq. (cm$^{-1}$)} && &\\
\cline{1-2}
Exp.~\cite{10}  & Calc.  && Symmetry & Proposed assignment \\
\br
150 & 174 &&   A$_1$(TO+LO) +  B$_1$(TO+LO)              & $R_c$(O'$_1$, O''$_1$) + $\tau$(O'$_2$, O''$_2$) + $t_b$(W)           \\
213 & 212 &&   B$_1$(TO+LO)                                          & $\delta$(O'$_3$, O''$_3$) + $t_b$(O'$_2$, O''$_2$) + $\tau$(O'$_1$, O''$_1$) + $t_b$(W)   \\
226 & 232 &&   A$_2$                                                       & $\delta$(O'$_2$, O''$_2$)   \\
265 & 255 &&  A$_1$(TO)                                                 & $\nu_a$(O'$_3$, O''$_3$) + $\delta$(O'$_2$, O''$_2$) + $R_c$(O'$_1$, O''$_1$)   \\
286 & 287 &&  B$_1$(TO) + A$_2$                                    & $\nu_a$(O'$_1$, O''$_1$) + $R_a$(O'$_2$, O''$_2$)   \\
311 & 300 &&  B$_2$(TO+LO) + A$_2$ + A$_1$(LO)         & $\nu_a$(O'$_3$, O''$_3$) + $R_b$(O'$_2$, O''$_2$)   \\
340 & 310 &&  B$_2$(TO+LO) + A$_2$ + A$_1$(TO)         & $\nu$(O'$_1$) + $\delta$(O''$_1$)   \\
377 & 364 &&  B$_2$(TO)                                                 & $\delta$(O'$_1$, O''$_1$)   \\
421 & 428 &&  A$_1$(LO) + B$_1$(TO+LO)                      & $\tau$(O'$_2$) + $\delta$(O'$_1$)  \\   
703 & 686 &&  B$_1$(TO) + A$_2$                                   & $\nu_a$(O'$_1$, O''$_1$)   \\ 
724 & 713 &&  A$_1$(TO) + B$_2$(TO)                            & $\nu_a$(O'$_1$, O''$_1$) + $\delta$(O'$_2$, O''$_2$)   \\ 
790 & 793 &&  B$_2$(TO+LO) + A$_1$(TO+LO)               & $\nu_a$(O'$_1$, O''$_1$) + $\delta$(O'$_2$, O''$_2$)   \\ 
827 & 814 &&  A$_1$(TO) + B$_2$(TO+LO)                     &$\nu_a$(O'$_1$, O''$_1$)   \\ 
842 & 829 && A$_1$(LO)                                                 &  $\nu_a$(O'$_1$, O''$_1$) \\ 
\br
\end{tabular}
\end{center}
\end{table}
\end{landscape}

%%%%%%%%%%%%%%%%%%%%%%%%%%%%%%%%%%%%%%%%%%%%%%%%%%%%%%%%%%%%%%%%%%%%%%%%%%
\section{Elastic and piezoelectric properties}
%%%%%%%%%%%%%%%%%%%%%%%%%%%%%%%%%%%%%%%%%%%%%%%%%%%%%%%%%%%%%%%%%%%%%%%%%%

Table~\ref{tab_el} reports the calculated single-crystal elastic $C_{ij}$ and compliances $S_{ij}$ tensors (relaxed-ion tensors~\cite{29}) of the Bi$_2$WO$_6$ in its ground-state ferroelectric structure. 
This structure properly appears mechanically stable since the elastic tensor constants satisfy the Born mechanical stability restrictions for an orthorhombic structure, which are given by the following system of inequations~\cite{33}:

\begin{equation}\label{1}
\left\{
\begin{array}{l}
C_{11}+C_{22} > 2C_{12} \\
C_{22}+C_{33} > 2C_{23} \\
C_{11}+C_{33} > 2C_{13} \\
C_{11}+C_{22}+C_{33}+2C_{12}+2C_{23}+2C_{13}>0 \\
C_{ii}>0,\ i=1,2,...,6.
\end{array}
\right.\end{equation}

By treating the polycrystalline materials as aggregates of single-crystals with random orientations, the isotropic polycrystalline elastic modulus can be computed as averages of anisotropic single-crystal elastic constants. The theoretical lower and upper bounds to the polycrystalline bulk ($B$) and shear ($G$) moduli are given by Reuss (assuming uniform stress throughout a polycrystal) and Voigt (assuming uniform strain) as follows~\cite{34}:

\begin{equation}
B_R=\frac{1}{S_{11} + S_{22} + S_{33} + 2(S_{12} + S_{23} + S_{13})}
\end{equation}
\begin{equation}
B_V=\frac{C_{11} + C_{22} + C_{33} + 2(C_{12} + C_{23} + C_{13})}{9}
\end{equation}
\begin{equation}
G_R=\frac{15}{4(S_{11} + S_{22} + S_{33}-S_{12}-S_{23}-S_{13} ) + 3(S_{44} + S_{55} + S_{66})}
\end{equation}
\begin{equation}
G_V=\frac{C_{11} + C_{22} + C_{33}-C_{12}-C_{23}-C_{13}}{15}+\frac{C_{44} + C_{55} + C_{66}}{5}
\end{equation}
 The polycrystalline values from the Voigt-Reuss-Hill approximation: $B = ( B_R + B_V )/2$ and $G = (G_R + G_V )/ 2$ are estimated using the different elastic parameters reported in Table~\ref{tab_el}.  The knowledge of these moduli, allow us to estimate the Young modulus ($Y$) and Poisson ratio ($\nu$) according to: $Y = 9 BG /(3B + G )$ and $\nu = (3B/2 - G )/ (3B + G )$. All these quantities are reported in Table~\ref{tab_el}. It is usually admitted that the typical relations between $B$ and $G$ are $G\simeq 1.1 B$ and $G\simeq 0.6 B$ for covalent and ionic materials, respectively. Thus, our calculations  indicate that the Bi$_2$WO$_6$ ferroelectric phase has a mixed ionic-covalent character ($G/B \approx 0.8$). This result is consistent with our $Z^\ast$ analysis.

\begin{table}[]
\caption{\label{tab_el} Elements of the elastic $C$ (GPa) and compliance $S$ (10$^{-2}$GPa$^{-1}$) tensors of Bi$_2$WO$_6$ in its $P2_1ab$ ferroelectric phase (relaxed-ion values \cite{29}). Voigt notation is used.}
\begin{center}
\begin{tabular}{lccccccccc}
\br
Index     & 11 &12  &13  &22  &23  &33  &44  &55  &66 \\
\br
$C$  & 167  &18 &44  &155 &52  &139  &69  &60  &68 \\
$S$  &0.65  & -0.01  & -0.20  & 0.73  & -0.27  & 0.88  & 1.44  & 1.66  & 1.46 \\
\br
\end{tabular}
\end{center}
\end{table}

\begin{table}[]
\caption{\label{tab_moduli} Bulk ($B$), Young ($Y$), shear ($G$) moduli, $G/B$ ratio and Poisson ratio ($\nu$) of Bi$_2$WO$_6$ polycrystalline  samples calculated according to Voigt, Reuss and Voigt-Reuss-Hill approximations.}
\begin{center}
\begin{tabular}{lccccccccc}
\br
                       & $B$ (GPa) & $G$ (GPa) & $G/B$& $Y$ (GPa) & $\nu$ \\
\br
Voigt                  &  76.55         &  62.57       & 0.81     & 147.51          & 0.17     \\
Reuss                  & 76.80       &  65.57       &   0.85   &   153.13        & 0.16      \\
Voigt-Reuss-Hill       & 76.67           &64.07           &  0.83   & 149.80          & 0.17      \\
\br
\end{tabular}
\end{center}
\end{table}

\begin{table}[]
\caption{\label{tab_piezo} Piezoelectric tensors (using Voigt notation) of Bi$_2$WO$_6$ in its $P2_1ab$ ferroelectric phase. The proper parts of the clamped-ion $\bar{e}$  and relaxed ion $e$ tensors are reported (in $C/m^2$) together with the relaxed ion $d$ tensor ($pC/N$) \cite{29}. Experimental data of the $d$ tensor ($d^{exp}$) are from Yanovskii {\it et al.}~\cite{8}.}
\begin{center}
\begin{tabular}{lccccc}
\br
Index     & 11 &12  &13  &26  &35 \\
\br
$\bar{e}$  & -0.21  &0.04  &0.23  &0.18 &-0.02   \\
$e$  & 3.45  &2.27  &1.14  &-0.60 &3.21   \\
$d$  &19.98  & 13.26  & -3.20  & -10.01  & 47.31   \\
$d^{exp}$  &40  &14   &-9 &-- &-- \\
\br
\end{tabular}
\end{center}
\end{table}

The piezoelectric $e$ (and $d$) tensor of the $P2_1ab$ ferroelectric phase of Bi$_2$WO$_6$ has five independent elements and writes :
\begin{equation}
e_{ij} =
\left (
\begin{array}{cccccc}
  e_{11}  &  e_{12}  &  e_{13}  &  \cdot  &  \cdot  & \cdot \\
  \cdot  &  \cdot  &  \cdot  & \cdot&  \cdot      &  e_{26}  \\
  \cdot  &  \cdot  &  \cdot  &  \cdot  &  e_{35}   &  \cdot  \\
\end{array}
\right ),
\end{equation}
The values of the proper parts of the clamped-ion $\bar{e}$  and relaxed ion $e$ tensors are reported in Table \ref{tab_piezo} together with the relaxed ion $d$ tensor \cite{29}. We first notice that the clamped-ion tensor $\bar{e}$ is very small  so that most of the piezoelectric response is produced by the ionic contribution; this contrasts with many other cases where both contributions are typically more comparable in amplitude and opposite in sign and partly cancel out \cite{35}. The agreement with the experiment is not particularly good except for $d_{12}$ but we notice that our values restrict to zero Kelvin. The signs are properly reproduced confirming that the crystal was oriented according to the same piezoelectric standards.

The experiment of Yanovskii {\it et al.}~\cite{8} suggests a sizable but not exceptionally large piezoelectric constants for this compound at room temperature, similarly to other aurivillius phases where relatively small piezoelectric responses were observed (less than 20 pC/N \cite{36}). These results contrasts with the cases of BaTiO$_3$ or PbZr$_{0.52}$Ti$_{0.48}$O$_3$ for which piezoelectric constants can reach 200~pC/N~\cite{36, 37}.  Now, at the level of the calculated intrinsic response at zero Kelvin, we notice that what is computed for Bi$_2$WO$_6$ is comparable to what was previously reported for PbTiO$_3$ and  PbZr$_{0.5}$Ti$_{0.5}$O$_3$ \cite{35}.

%%%%%%%%%%%%%%%%%%%%%%%%%%%%%%%%%%%%%%%%%%%%%%%%%%%%%%%%%%%%%
\section{Second-order nonlinear optical susceptibility}
%%%%%%%%%%%%%%%%%%%%%%%%%%%%%%%%%%%%%%%%%%%%%%%%%%%%%%%%%%%%%

The second-order nonlinear optical susceptibility, $\chi^{(2)}$, is a third-rank tensor related to the electronic response of the system and depends on the frequencies of the optical electric fields~\cite{38}. However, in the present context of the $2n + 1$ theorem applied within the LDA to (static) DFT, we neglect the dispersion of $\chi^{(2)}$ computing the electronic response at zero frequency. Within these conditions, the $\chi^{(2)}$-tensor is related to a third-order derivative of a field-dependent energy functional, ${\cal F}=E-\Omega_0 \bm{\mathcal{E.P}}$, where $E$ and ${\bm{\mathcal{P}}}$ are respectively the total energy in zero field and the macroscopic polarization~\cite{18}. As a consequence, the $\chi^{(2)}$-tensor satisfies Kleinman's symmetry condition~\cite{39} and its indices are therefore symmetric under a permutation. As usual in nonlinear optics, we report the so-called ``$d$-tensor'', defined as $ d^{NLO} = \frac{1}{2} \chi^{(2)}$ (a superscript ``NLO'' is introduced here to avoid confusion with the piezoelectric tensor).

In the $P2_1ab$ ferroelectric phase, using Kleinman's symmetry rule allows to reduce the number of independent elements of the non-linear optic $d^{NLO}$-tensor to three  ( $d^{NLO}_{26} = d^{NLO}_{12}$ and $d^{NLO}_{35} = d^{NLO}_{13}$) so that it writes:

\begin{equation}
d^{NLO}_{ij} =
\left (
\begin{array}{cccccc}
  d^{NLO}_{11}  &  d^{NLO}_{12}  &  d^{NLO}_{13}  &  \cdot  &  \cdot  & \cdot \\
  \cdot  &  \cdot  &  \cdot  & \cdot&  \cdot      &  d^{NLO}_{12}  \\
  \cdot  &  \cdot  &  \cdot  &  \cdot  &  d^{NLO}_{13}   &  \cdot  \\
\end{array}
\right ),
\end{equation}
where the indices $i$ and $j$ denote the Cartesian components in Voigt notation. The calculated values are given in Table~\ref{tab_dtensor} in comparison with the available experimental data (the d$^{NLO}_{12}$ element is still unmeasured). 

We observe a reasonably good agreement between the theoretical and experimental absolute values. We notice that although all calculated susceptibilities are found to be negative, positive values have been reported experimentally. It is not clear however if the signs have been unambiguously determined experimentally;  they depend on the direction of the axis and sign disagreements are often observed between authors on ferroelectric materials~\cite{24,25}. The magnitudes of the calculated $d^{NLO}_{11}$ and d$^{NLO}_{12}$ coefficients are very similar, whereas $d^{NLO}_{13}$ is significantly smaller, in agreement with the experiment. We notice that the scissors correction decreases significantly the values of the nonlinear optical susceptibilities, but does not change this trend. It seems that the scissors correction leads to a more important discrepancy with the experimental results. As discussed in Ref.  \cite{18}, this is not necessarily unusual  since various errors (like the LDA volume underestimate) can also significantly affect the values of $d^{NLO}_{ij}$. The amplitude of the scissors correction can be considered as a rough estimate of the theoretical uncertainty.  
\begin{table}[h]
\begin{center}

\caption{\label{tab_dtensor}Nonlinear optical susceptibility values of Bi$_2$WO$_6$ in its $P2_1ab$ ferroelectric phase. The effect of scissors correction (0.7 eV) is also reported.}
\begin{tabular}{lcccccc}
\br
&d$^{NLO}_{11}$ &d$^{NLO}_{12}$ &d$^{NLO}_{13}$  \\
\hline
LDA&-9.57 & -8.05&-4.55 \\
LDA+Sciss.&-6.42 & -5.48&-3.37 \\
Exp.~\cite{8}& 8.2 &-- & 5.0\\
\br
\end{tabular}
\end{center}
\end{table} 

%%%%%%%%%%%%%%%%%%%%%%%%%%%%%%%%%%%%%%%%%%%%%%%%%%%%%%%%%%%%%%%%%%%%%%%%%%%%%%%%%%%%%%%%%%%%%%%%%%%%
\section{Pockels response}
%%%%%%%%%%%%%%%%%%%%%%%%%%%%%%%%%%%%%%%%%%%%%%%%%%%%%%%%%%%%%%%%%%%%%%%%%%%%%%%%%%%%%%%%%%%%%%%%%%%%

Similarly to the second-order nonlinear optical susceptibility, electro-optic (EO) tensor (Pockels response) has five independent coefficients given by:
\begin{equation}
r_{ij} =
\left (
\begin{array}{ccc}
  r_{11}   & \cdot   &  \cdot   \\
 r_{21}   & \cdot   &  \cdot   \\
 r_{31}   & \cdot   &  \cdot   \\
 \cdot   & \cdot   &  \cdot   \\
 \cdot   & \cdot   &  r_{53}   \\
 \cdot   & r_{62}   &  \cdot   \\
\end{array}
\right ).
\end{equation}

We restrict ourselves to the computation of the so-called ``clamped'' EO coefficients, neglecting any modification of the unit cell shape due to the inverse piezoelectric effect. Within the Born-Oppenheimer approximation, the clamped EO tensor can be decomposed as the sum of two terms: an electronic term and an ionic term \cite{18,19}. The electronic term describes the interaction of the quasi-static electric field with the valence electrons when considering the ions artificially clamped at their equilibrium positions. It is proportional to the second-order nonlinear optical susceptibility. The ionic term takes into account the relaxation of the atomic positions in the applied quasi-static electric field and the variations of $\varepsilon_{ij}$ induced by these displacements. It is proportional to the Raman susceptibilities, the mode polarities and to the inverse square of the frequencies~\cite{18, 38}. The $r_{62}$ and $r_{53}$ ionic terms are respectively linked to the B$_1$ and B$_2$-modes, whereas the three $r_{11}$, $r_{21}$ and $r_{31}$ ionic terms are linked to the A$_1$-modes. The EO coefficients are found relatively small and not exceptional despite the presence of very low frequencies modes that could have suggested higher EO coefficients. This is due to the small Raman susceptibilities of these low frequencies modes and to their small polarities. The mode-by-mode decomposition of each TO phonons to the ionic terms is also reported in Table~\ref{tab_EO}. 

\begin{table}[tbp]
\caption{\label{tab_EO} Calculated electronic and ionic contributions of the TO phonon modes to the clamped EO tensor (in pm/V) of Bi$_2$WO$_6$ in its $P2_1ab$ ferroelectric phase. Frequencies, $\omega_m$, are given in cm$^{-1}$. Coefficients obtained using scissors corrected values for the refractive indices are also reported for comparison.}
\begin{center}
\begin{tabular}{lrccccccccccc}
\br
&&&\multicolumn{4}{c}{A$_1$-modes}&&\multicolumn{2}{c}{B$_2$-modes}&&\multicolumn{2}{c}{B$_1$-modes}\\
\cline{4-7}\cline{9-10}\cline{12-13}
          &     && $\omega_m$ & $r_{11}$ & $r_{21}$ & $r_{31}$ && $\omega_m$ & $r_{53}$ && $\omega_m$ & $r_{62}$  \\
\hline
Electronic (LDA)&     &&     &  0.88 &  0.72 &  0.42 &&     &  0.42 &&     & 0.73   \\
\hline
Ionic (LDA)     &  &&  35 &  0.02 &  0.02 & -0.03 &&  50 &  0.02 &&  36 &  0.16  \\
          & &&  62 &  0.73 &  1.13 & -0.02 &&  75 & -0.01 &&  59 &  0.13   \\
          &  &&  75 & -0.27 &  0.10 &  0.11 &&  78 & -1.12 &&  87 & -0.02   \\
          &  &&  84 & -0.08 & -0.01 &  0.00 &&  82 & 0.00 && 103 &  0.01   \\
          &  &&  96 & -0.02 & -0.01 & -0.01 &&  96 &  1.36 && 112 &  0.31   \\
          &  && 114 & -0.11 &  0.10 &  0.20 && 100 & 0.00 && 166 & 0.00   \\
          &  && 166 &  2.31 &  2.31 &  0.13 && 161 &  0.05 && 167 &  0.00   \\
          &  && 175 & -0.14 & -0.11 & -0.13 && 166 &  0.00 && 174 &  0.03    \\
          &  && 181 &  0.02 &  0.02 &  0.00 && 168 &  0.03 && 191 &  0.00  \\
          & && 186 & -0.01 &  0.08 &  0.16 && 185 &  0.01 && 212 &  4.04 \\
          & && 254 &  2.07 &  2.75 &  2.01 && 268 &  0.01 && 268 &  0.01  \\
          & && 262 &  0.06 &  0.04 &  0.05 && 275 & -0.06 && 289 &  0.55 \\
          & && 268 &  0.12 &  0.12 &  0.08 && 291 & -0.03 && 311 &  0.58 \\
          & && 274 &  0.63 &  1.02 &  0.72 && 297 &  0.02 && 331 & -0.01 \\
          & && 283 &  0.05 &  0.09 &  0.04 && 300 & -0.39 && 334 &  0.05 \\
          & && 309 &  1.29 &  0.35 &  0.10 && 309 &  0.00 && 341 &  0.00 \\
          & && 353 & -0.08 & -0.21 & -0.10 && 365 &  0.90 && 418 &  0.00 \\
          & && 358 & -0.01 & -0.02 & -0.01 && 425 &  0.03 && 429 & -0.07 \\
          & && 386 & -0.02 & -0.02 & -0.03 && 453 &  0.02 && 447 & -0.04 \\
          & && 390 & 0.00 & 0.00 & -0.01 && 491 &  0.00 && 450 & -0.05 \\
          & && 422 &  0.01 &  0.00 & 0.00 && 519 &  0.04 && 543 &  0.01 \\
          & && 428 &  0.00 & -0.01 &  0.01 && 585 &  0.00 && 557 &  0.01 \\
          & && 713 &  1.10 &  0.91 &  0.41 && 694 & -0.03 && 634 &  0.04 \\
          & && 776 &  0.00 &  0.01 &  0.00 && 714 &  0.00 && 686 &  0.85 \\
          & && 793 & -0.01 & -0.01 & -0.01 && 793 &  0.00 && 708 &  0.02 \\
          & && 814 &  0.07 &  0.09 &  0.04 && 811 &  0.00 && 998 &  0.00 \\
\multicolumn{2}{r}{Sum of phonons} &&&7.74 & 8.73 & 3.73 &&  & 0.82  &&&  6.69 \\
\hline
Total (LDA)     &     &&     &  8.61 &  9.45 &  4.15 &&     & 1.24  &&     &  7.34 \\
\hline
Total (LDA + SCI)     &     &&     &  10.11 &  11.13 &  4.88 &&     & 1.46  &&     &  8.63 \\
\br
Exp.~\cite{8}         &              && & 19  & 13 & 15          &&         &--       &&       &--\\
\br
\end{tabular}
\end{center}
\end{table}

We observe that the electronic and the total ionic terms show a significant anisotropy, yielding to anisotropic  total EO coefficients. These two terms have the same sign (positive) for all nonzero coefficients. So, they act constructively to increase the global EO coefficients. Nevertheless, the electronic and the total ionic terms do not equally contribute to the total EO coefficients. Except for the $r_{53}$ coefficient, the electronic terms contribute by around 10\% to each total EO coefficient whereas the ionic terms contribute by around 90\%. Indeed, the $r_{62}$ ionic term is mainly dominated by four B$_1$-modes: B$_1$(212 cm$^{-1}$), B$_1$(289 cm$^{-1}$), B$_1$(311 cm$^{-1}$) and B$_1$(686 cm$^{-1}$) and they contribute by around 55, 7, 8 and 12\% to the $r_{62}$ total EO coefficient. The same sign (positive) of these four modes therefore leads to a significant value of the total $r_{62}$ ionic term. The same explanation can be given for the significant contribution of the $r_{11}$, $r_{21}$ and $r_{31}$ ionic terms to the corresponding total EO coefficients. In the case of the $r_{53}$ coefficient, its ionic term is mainly dominated by four B$_2$-modes: B$_2$(78 cm$^{-1}$), B$_2$(96 cm$^{-1}$), B$_2$(300 cm$^{-1}$) and B$_2$(365 cm$^{-1}$). However, these modes are compensated between them due to their similar magnitude and their different signs. As a consequence, the $r_{53}$ ionic contribution significantly decreases to 66\% while that of electronic term is increased to 34\%. 

A scissors correction can also be applied. As in Ref. \cite{18}, its effect was restricted to the refractive indices (see Table \ref{optdielc}) entering in the calculation of $r_{ij}$ (no ``scissors'' correction was considered for the non-linear optical susceptibilities since, as discussed in the previous Section, it does not seem to provide any improvement with respect to the experiment). This correction tends to increase the total EO coefficients as reported in Table~\ref{tab_EO}, but does not dramatically change their amplitude. Our calculated EO coefficients significantly underestimate the experimental values reported by Yanovskii {\it et al.} (at 1 kHz)~\cite{8}. This can eventually be related to different effects: (i) we computed the ``clamped'' EO coefficients while the strain relaxation could significantly affect the results through the piezoelectric effect \cite{20} (ii) we worked at 0~K while the influence of the temperature should be considered when comparing our EO predictions to the experimental ones \cite{40}).

%%%%%%%%%%%%%%%%%%%%%%%%%%%%%%%%%%%%%%%
\section{Conclusions}
%%%%%%%%%%%%%%%%%%%%%%%%%%%%%%%%%%%%%%%

In this work, we have performed a systematic  first-principles study of Bi$_2$WO$_6$ in its $P21ab$ ferroelectric ground state using density functional theory  within the local density approximation. We have investigated important physical properties like dielectric, dynamical, elastic and piezoelectric responses,  second-order nonlinear optical susceptibilities and Pockels coefficients. Infrared and Raman spectra have been calculated. In spite of the large complex structure of Bi$_2$WO$_6$, the agreement between experimental and calculated spectra is good enough to provide reliable assignments of the observed infrared and Raman bands.  As in simple perovskite ferroelectric oxides, the Born effective charges are anomalous. A sizable intrinsic piezoelectric effect is reported and the elastic properties show a mixed ionic-covalent character, in agreement with the analysis of Born effective charges. Finally, we also calculated the nonlinear optical susceptibilities and the ``clamped'' electro-optic coefficients, quantifying its purely electronic and ionic contributions. We found that Bi$_2$WO$_6$ exhibits various sizable functional properties, although no exceptionally large values have been identified in comparison to simple ferroelectric perovskite compounds. Our work reports benchmark theoretical data, usually in good qualitative agreement with the experimental work of Maczka {\it et al.} \cite{10}. To date, this compound was however only poorly characterized and we hope that this work will motivate further experimental measurements.

%%%%%%%%%%%%%%%%%%%%%%%%%%%%%%%%%%%%%%%
\section*{Acknowledgments}
%%%%%%%%%%%%%%%%%%%%%%%%%%%%%%%%%%%%%%%
This work was performed during a visit of H.D. at the University of Li\`ege (Belgium) with the combined support of the Algerian Ministry of High Education and Scientific Research (MESRS) and the InterUniversity Attractive Pole Program from the Federal Science Research Policy of Belgium. Ph.G. thanks the Francqui Foundation for Research Professorship and the ARC project The MoTherm. The authors thank Dr. M. Maczka, for sending his experimental data.
\\
%%%%%%%%%%%%%%%%%%%%%%%%%%%%%%%%%%%%%%%%%%%%%%%%%%%%%%%%%%%%%%%%%%%%%%%%%%%%%%%%%%%%%%%%%%%%%%%%%%%%%%%%%%%%%%%%%%%%%%%
%%%%%%%%%%%%%%%%%%%%%%%%%%%%%%%%%%%%%%%%%%%%%%%%%%%%%%%%%%%%%%%%%%%%%%%%%%%%%%%%%%%%%%%%%%%%%%%%%%%%%%%%%%%%%%%%%%%%%%%


\begin{thebibliography}{40}

\bibitem{1}
R. H. Mitchell, ``Perovskites: modern and ancient'' (Almaz Press Inc., Thunder Bay, 2002). 
\bibitem{2}
B. Aurivillius, {Ark. Kemmi},{1}, {463}, {1949}.
\bibitem{3}
Y. Li, J. Liu, X. Huang and G. Li, {Cryst. Growth Des.}, {7}, {1350}, {2007}.
\bibitem{4}
N. Kim, R. N. Vannier and C. P. Grey , {Chem. Mater.}, {17}, {1952}, {2005}.
\bibitem{5}
A. Kudo and S. Hijii, {Chem. Lett.}, {1103}, {1999}.
\bibitem{6}
N. A. McDowell, K. S. Knight and P. Lightfoot, {Chem. Eur. J.}, {12}, {1493}, {2006}.
\bibitem{7}
H. Djani, E. Bousquet, A. Kellou and Ph. Ghosez, {Phys. Rev. B}, {86}, {054107}, {2012}.
\bibitem{8}
V. K. Yanovskii and V. I. Voronkova, {Phys. Stat. Sol. (a)}, {93}, {57}, {1986}.
\bibitem{9}
M. Maczka, J. Hanuza, W. Paraguassu, A.G. Souza Filho, P. T. C. freire and J. Mendes Filho, {Appl. Phys. Lett.}, {92}, {112911}, {2008}.
\bibitem{10}
M. Maczka, L. Macalik, K. Hermanowicz, L. Kepinski and P. Tomaszewski, {J. Raman Spectrosc.}, {41}, {1059}, {2010}.
\bibitem{11} 
X. Gonze, B. Amadon, P.-M. Anglade, J.-M. Beuken, F. Bottin, P. Boulanger, F. Bruneval, D. Caliste, R. Caracas, M. Cote, T. Deutsch, L. Genovese, Ph. Ghosez, M. Giantomassi, S. Goedecker, D.R. Hamann, P. Hermet, F. Jollet, G. Jomard, S. Leroux, M. Mancini, S. Mazevet, M. J. T. Oliveira, G. Onida, Y. Pouillon, T. Rangel, G.-M. Rignanese, D. Sangalli, R. Shaltaf, M. Torrent, M. J. Verstraete, G. Zerah and J. W. Zwanziger, {Computer Phys. Comm.}, {180}, {2582}, {2009}.
\bibitem{12} 
X. Gonze, J.-M. Beuken, R. Caracas, F. Detraux, M. Fuchs, G.-M. Rignanese, L. Sindic, M. Verstraete, G. Zerah, F. Jollet, M. Torrent, A. Roy, M. Mikami, Ph. Ghosez, J.-Y. Raty, D.C. Allan, {Comput. Mater. Sci.}, {25}, {478}, {2002}.
\bibitem{13} 
X. Gonze, G.-M. Rignanese, M. Verstraete, J.-M. Beuken, Y. Pouillon, R. Caracas, F. Jollet, M. Torrent, G. Zerah, M. Mikami, Ph. Ghosez, M. Veithen, V. Olevano, L. Reining, R. Godby, G. Onida, D. Hamann, D. C. Allan, {Zeit. Kristallogr.}, {220}, {558}, {2005}. 
\bibitem{14}
M. P. Teter, {Phys. Rev. B}, {48}, {5031}, {1993}. 
\bibitem{15}
H. J. Monkhorst and J. D. Pack, {Phys. Rev. B}, {13}, {5188}, {1979}.
\bibitem{16}
H. B. Schlegel, {J. Comp. Chem.}, {3}, {214}, {1982}.
\bibitem{17}
X. Gonze and C. Lee, {Phys. Rev. B}, {55}, {125107}, {1997}.
\bibitem{18}
M. Veithen, X. Gonze, Ph. Ghosez, {Phys. Rev. B}, {71}, {125107}, {2005}.
\bibitem{19}
M. Veithen, X. Gonze, Ph. Ghosez, {Phys. Rev. Lett.}, {93}, {187401}, {2004}.



\bibitem{20} 
P. Hermet, J. L. Bantignies, J. L. Sauvajol, and M. R. Johnson, {Synth. Met.} {156}, {519}, {2006}.
\bibitem{21} 
P. Hermet,  L. Gourrier, J.-L. Bantignies, D. Ravot, T. Michel, S. Deabate, P. Boulet, F. Henn, {Phys. Rev. B}, {84}, {235211}, {2011}.
\bibitem{22}
Ph. Ghosez, J.P. Michenaud and X. Gonze, {Phys. Rev. B}, {58}, {6224}, {1998}. 
\bibitem{23}
W.A. Harrison  {Electronic Structure and the Properties of Solids}, {San Fransisco, CA: Freeman}, {1980}.
\bibitem{24}
P. Hermet, M. Veithen and Ph. Ghosez {J. Phys. Condens. Matter}, {19}, {456202}, {2007}.
\bibitem{25}
P. Hermet, M. Veithen and Ph. Ghosez {J. Phys. Condens. Matter}, {21}, {215901}, {2009}. 
\bibitem{26} 
X. Gonze, Ph. Ghosez and R. W. Godby, {Phys. Rev. Lett}, {74}, {4035},  {1995}.
\bibitem{27}
Z. H. Levine  and D. C. Allan, {Phys. Rev. Lett.}, {63}, {1719}, {1989}.
\bibitem{28}
A. Kudo and  S. Hijii, {Chem. Lett. }, {10}, {1103}, {1999}.
\bibitem{29}
X. Wu, D. Vanderbilt, D. R. Hamman, {Phys. Rev. B}, {72}, {035105}, {2005}.

\bibitem{30}
P. Hermet, M. Goffinet, J. Kreisel, Ph. Ghosez, {Phys. Rev. B}, {75}, {220102(R)}, {2007}.
\bibitem{31}
S. A. Prosandeev, U. Waghmare, I. Levin, J. Maslar, {Phys. Rev. B}, {71}, {214307}, {2005}.
\bibitem{32}
R. Caracas, E. J. Banigan, {Chem. Phys.}, {127}, {144510}, {2007}.

\bibitem{33}
D. C. Wallace, {Thermodynamics of Crystals}, {Wiley, New York}, {1972}.
\bibitem{34}
K. B. Panda and K. S. R. Chandran, {Acta Mater.}, {54}, {1641}, {2006}.
\bibitem{35}
L. Bellaiche and D. Vanderbilt, {Phys. Rev. Lett.}, {83}, {1347}, {1999}.
\bibitem{36}
M. Zgonik, P. Bernasconi, M. Duelli, R. Schlesser, P. G\"{u}nter, M.H. Garett, D. Rytz, Y. Zhu, X. Wu, {Phys. Rev. B}, {50}, {5941}, {1994}.
\bibitem{37}
B. Jaffe, W.R. Cook, H. Jaffe, {Piezoelectric Ceramics}, {3}, {Academic press Inc. New york}, {1971}.
\bibitem{38} 
J. L. P. Hughes and , J. E. Sipe, {Phys. Rev. B} {53}, {10751}, {1996}.
\bibitem{39} 
D. A. Kleinman, {Phys. Rev.}, {126}, {1977}, {1962}.
\bibitem{40}
M. Veithen and Ph. Ghosez, {Phys. Rev. B} {71}, {132101}, {2005}.

\end{thebibliography}
\end{document}